\documentclass[12pt,a4paper]{article}

\usepackage[font=scriptsize]{caption}
\usepackage{titlesec}
\usepackage{multicol}
\usepackage{flushend}
\usepackage{amsfonts}
\usepackage{setspace}
\usepackage{balance}
\usepackage{nomencl}
\usepackage{graphicx}
\usepackage[numbers]{natbib}
\usepackage{amsmath}
\usepackage{amsthm}
\usepackage{amsmath}
\usepackage{amssymb}
\usepackage{stmaryrd}
\usepackage{stackrel}
\usepackage{multicol}
\usepackage{url}
\usepackage{color,hyperref}
\usepackage{subfig}

\begin{document}



\title{The null geodesics of charged and non--charged black hole in mimetic gravity}

\maketitle


\begin{center}
\author{{M. Haditale}*, {B. Malekolkalami}}

\thanks{Faculty of Science, University of Kurdistan, Sanandaj, P. O. Box 416, Iran (email:
Maryam.haditale@uok.ac.ir, B.Malakolkalami@uok.ac.ir)}
\end{center}

\begin{abstract}
The null geodesics around the charged black hole spacetimes are investigated in the mimetic gravity framework when Einstein's gravity is coupled to a nonlinear electromagnetic field.  The photon paths in nonlinear electrodynamics are geodesics of the effective metric which is determined by the background metric and the particular nonlinear theory considered.
The nonlinear effects are represented by a quadruple moment and appear as a correction term to Reissner Nordstrom (\textbf{\emph{RN}}) metric and Reissner Nordstrom-anti-de Sitter (\textbf{\emph{RN--(A)dS}}) in cylindrical metric. Remarkably, the nonlinear effects prevent the circular orbits around black holes, and the presence of the mimetic field can change the repulsive character of dS space to the attractive one. Also, the effect of the mimetic parameter manifests itself in the stronger or weaker gravity of the black hole.
\end{abstract}
\section{Introduction}\label{sec-1}
General Relativity (\textbf{\emph{GR}}) is the geometry theory of gravitation proposed by Einstein to describe gravitation in modern physics. One of the most interesting predictions of the GR is Black Hole (\textbf{\emph{BH}}) \cite{Airy1} as a super condensed mass that distorts the spacetime around itself. \emph{BH} is a region of spacetime where gravity is so strong that nothing no particles or even light radiation can escape from it. The boundary of such a region is called the event horizon.
Despite obtaining different solutions for charged and non-charged \emph{BHs} in the \emph{GR} theory, this theory is problematic in the singularity points and hence quantum effects must be taken into account \cite {Airy2}. To avoid the singularity problem, theories such as mimetic gravity and dark matter have been proposed \cite {Airy3}. Many changes of original mimetic gravity are defined in the literature \cite {Airy4}. The mimetic gravity theories are used to discuss the caustic and ghost instabilities and the cosmological evolutions \cite{Airy5, Airy6}.\\

Many modified mimetic theories have been considered to the quantum corrections: mimetic $F(R)$ \cite{Airy7}, $F(R,T)$ \cite{Airy8} and $F(R,\phi )$ \cite{Airy19}, mimetic covariant Horava--like gravity \cite{Airy10}, mimetic Galileon gravity \cite{Airy11}, mimetic Born--Infeld gravity \cite{Airy12}, mimetic Horndeski gravity \cite{Airy13}, unimodular--mimetic $F(R)$ gravity \cite{Airy14}, non-local mimetic $F(R)$ gravity \cite{Airy15}, the vector--tensor gravity \cite{Airy16} and bi--scalar mimetic gravity \cite{Airy17} are some examples in this field.\\

On the other hand, the coupling between gravitational theories and non--linear electrodynamics can give a background to fix the problems in cosmology and astrophysics. For example it can provide the necessary pressure for the expansion of the universe, equivalently, non--linear electrodynamics plays the role of dark energy \cite {Airy18, Airy19}. Also, the coupling of nonlinear electrodynamic with gravity results effective models taking into account loop quantum correction to Maxwell electrodynamic. Such models are usually utilized to eliminate the singularity of electromagnetic fields which are welcomed by many of people. From the pioneer works in this subject is the Born-Infeld theory \cite {Airy20}. The importance and successes of the coupling of gravity with nonlinear electrodynamics are described in detail in the references \cite {Airy21}.\\

In the present work, we focus on null geodesics around the charged and non--charged new \emph{BH} solutions in the framework of the mimetic gravity coupled to non--electrodynamic \cite{Airy22}. These new solutions are the result of coupling the mimetic gravity with non--electrodynamics and include quadruple moment which appears as a correction to \emph{RN} \emph{BH}s. The following points can be mentioned for the importance of such models:\\
1) The main role of the quadruple moments in waves radiation (including gravitational ones).\\
2) The gravity theories as \textbf{\emph{ABGB}}, e. g. \cite {Airy23} and quadratic gravity, e. g. \cite {Airy24} present the quadruple solution in appropriate limits.\\
3) The quadruple solution is related to a constant so that its vanishing makes the solutions coincide with the linear Maxwell \emph{BH}s (as stated  \cite {Airy22, Airy25}).\\

The importance of geodesics in this regard goes back to the successful applications of mimetic theory in astrophysics and cosmology. For instance, mimetic gravity provides a framework for understanding flat galactic rotation curves through the Schwarzschild (\textbf{\emph{Sch}}) modification of spacetime \cite {Airy26} and or its application in the solar system \cite {Airy27}- \cite{Airy32}.\\

In reference \cite{Airy22}, the author discusses some of the singularities and thermodynamic properties of such \emph{BH} solutions. An important feature of the solutions is asymptotically flat or \emph{(A)dS} and the charged solutions are considered in both linear and nonlinear regimes. Here, our interest is related to the study of light paths (null geodesics) in corresponding spacetimes for a model of nonlinear electrodynamics. Generally, geodesy is the discipline that deals with the measurement of free particle paths in any gravitational theory. They are one of the most key concepts related to the fundamental structure of any spacetime. Indeed, they are inertial trajectories of spacetime and so finding such paths is of particular importance. On other hand, seeking new \emph{BH} solutions is extremely relevant to set up any relativistic theory of gravity, especially in the case of \emph{(A)dS} \emph{BH} (e. g. \cite{Airy33}). Consequently, exploring the corresponding geodesics can be doubly important. In other words, to confirm the new solutions for any relativistic theory of gravitation,  the existence of geodesics (especially null geodesics) can be regarded as the main challenge. Especially, the trajectory of a photon is the key to understanding and exploring the physics of spacetime geometry and structure. \\

For this aim, task one is to define spacetime metric, but the paths of photons in nonlinear electrodynamics are not null geodesics of the background geometry. Instead, they follow null geodesics of an effective metric determined by the nonlinearities of the electromagnetic field, which depends on the particular nonlinear theory considered.

To be more precise, the authors are interested in studying the null geodesics around the charged \emph{BH} in presence of a mimetic field taking into account the nonlinearity effects. But, to make a comparison between the nonlinear and linear cases, a section of the work is dedicated to a review of the simplest charged and uncharged \emph{BH}. For a complete discussion about this subject, the reader can refer to the original texts, e. g. \cite{Airy1}.\\
The work is divided into the following sections:\\
In Sec \ref{sec-2}, a brief overview of the basic concepts for the work is given. For Sec. \ref{sec-3}, the null geodesics for the simplest (charged and non--charged \emph{BH}s) are presented in linear electrodynamics. In Sec. \ref{sec-4}, the null geodesics of \emph{BH} solutions in the nonlinear electrodynamic and the presence of mimetic gravity are investigated. The conclusions are given in Section \ref{sec-5}.

\section{The Background and  Effective Metrics}\label{sec-2}
The most general static and spherically symmetric line element in four dimensions (including charged objects) is of the form
\begin{equation}
{ds^2}=f (r)\mathop{dt^2}-\frac{\mathop{dr^2}}{f (r)}-\mathop{r^2}\mathop{d\Omega ^2},
\label{eq1}
\end{equation}
where the function $f(r)$ is determined by the corresponding field equations.\\
The interest in avoiding singularity and other related problems pushes people to the other \emph{BH}s theories or models \cite{Airy34}.
One of the known models for avoiding singularity in the presence of electric charge is the Einstein gravity coupled to Born--Infeld electrodynamics \cite{Airy20}. After this, Hoffman proposed to couple the General  Relativity with Born--Infeld electrodynamics to obtain a spherically symmetric solution representing the gravitational field of a charged object \cite{Airy35}.

To study the geodesic structure of massless
particles in nonlinear electrodynamics, the effective metric should be used instead of the background metric. In other words, if the (background) metric (1) describes  spacetime corresponding to Einstein gravity coupled to Born--Infeld electrodynamics, the effective metric for null geodesics becomes \cite{Airy36}:
\begin{equation}
{ds_{eff}^2}=\mathop{\omega^{1/2}(r)}f (r)\mathop{dt^2} -\frac{\mathop{\omega^{1/2}(r)}}{f (r)}\mathop{dr^2}-\mathop{\omega^{-1/2}(r)}\mathop{r^2}\mathop{d\Omega ^2},
\label{eq2}
\end{equation}
where the function $\omega (r)$  is determined by a particular nonlinear theory considered. It is obvious from (1) and (2) that, the horizon structure of the effective metric and background metric is the same, but the photon paths are different, because,  the effective metric contains function $\omega(r)$.

In this work, we are interested in studying the geodesic paths of the massless particle around spherical symmetric charged (and non--charged) \emph{BH} in Mimetic Gravitational Theory \cite{Airy22} and consequently the function $\omega(r)$ is determined by the corresponding field equations. The original formulation of the mimetic theory of gravity can be obtained starting from general relativity, by isolating the conformal degree of freedom of gravity by introducing a parametrization of the physical metric in terms of an auxiliary metric and a scalar field, dubbed mimetic field $\omega (r) $. The relation between these is:
\begin{equation}
\mathop{g_{\alpha \beta }}=\left(\mathop{{\bar{g}}^{\mu \nu }}\mathop{\partial _{\mu }}\omega \mathop{\partial _{\nu }}\omega\right)\mathop{{\bar{g}}_{\alpha \beta }},
\label{eq3}
\end{equation}
where $\mathop{{g}}_{\mu \nu}$, $\mathop{{\bar{g}}}_{\alpha \beta}$ are   physical  and auxiliary metrics, respectively. Also, from definition (3), it is easy to check that \cite{Airy22}:
\begin{equation}
\mathop{g^{\mu \nu }}\mathop{\partial _{\mu }}\omega \mathop{\partial _{\nu }}\omega =-1,
\label{eq4}
\end{equation}
which can be considered as a lateral constraint on the mimetic field determined by the field equations.

\section{The Background Metrics in the Linear Electrodynamic}\label{sec-3}
In this section, we first consider metric (1) for elementary and simplest charged and non--charged \emph{BH}s to explore the null geodesics in linear electrodynamics. Although some results may be repetitive (as done in the previous texts and works), they are given for comparison.

Remarkably, the auxiliary metric and the mimetic field don't appear in the field equations \cite{Airy22}, and to write down the geodesic equations, the background metric is sufficient. The geodesic equations are in the spherically symmetric and cylindrically spacetimes.

\subsection{Spherically Symmetric Metric}\label{subsec-1}
Three introductory \emph{BH}s considered are \emph{Sch}, \emph{RN}, and Schwarzschild--de Sitter (\textbf{\emph{$Sch-\Lambda$}}) ones which the corresponding background functions in (1) are given by:
\begin{equation}
\mathop{f_{Sch} (r)}=1-\frac{2M}{r}
\label{eq5}
\end{equation}
\begin{equation}
\mathop{f_{RN} (r)}=1-\frac{2M}{r}+\frac{\mathop{q}^{2}}{\mathop{r}^{2}}
\label{eq6}
\end{equation}
\begin{equation}
\mathop{f_{Sch-\Lambda }(r)}=1-\frac{\Lambda \mathop{r}^{2}}{6}-\frac{2M}{r},
\label{eq7}
\end{equation}
where $M$, $q$, and $\Lambda $ have usual meanings. The above background functions are plotted in Fig.1.\\
\subsection{Cylindrically  Symmetric Metric}\label{subsec-2}
\begin{figure}[!pt]
\centering
\includegraphics[width=14cm]{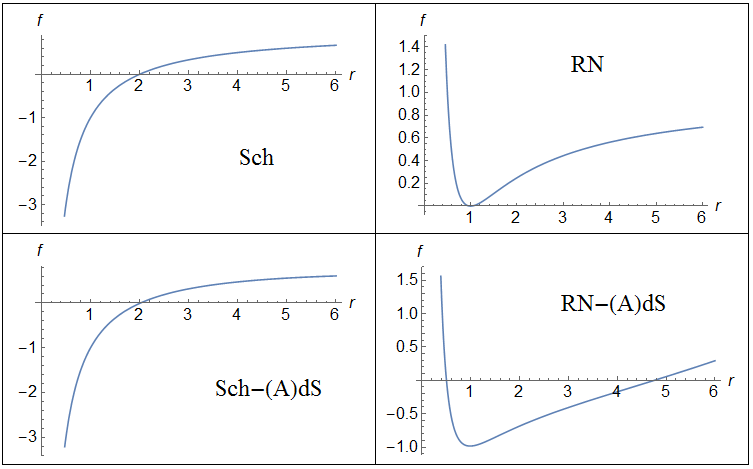}
\caption{The graph of the metric coefficient (1) in linear electrodynamics for the cases \emph{Sch} (5), \emph{RN} (6), \emph{Sch–(A)dS} (7) and \emph{RN-(A)dS} (9). Numerical values of parameters are: $M=1$, $q=1$ and $\Lambda=-0.9$.}
\label{fig1-1}
\end{figure}

Asymmetric spacetime metric in cylindrical coordinates $(r, \phi, z)$  can be written as:
\begin{equation}
\mathop{ds^{2}}=f(r)\mathop{dt^{2}}-\frac{1}{f(r)}\mathop{dr^{2}}-\mathop{r^{2}}(\mathop{d\phi ^{2}}+\mathop{dz^{2}}),
\label{eq8}
\end{equation}
where the background function $f(r)$ is determined by the corresponding field equations.
Here, we consider \emph{RN-(A)dS} spacetime in cylindrical coordinates which the corresponding background function is given by \cite{Airy22}:
\begin{equation}
\mathop{f_{RN-\Lambda }(r)}=\frac{\mathop{q}^{2}}{\mathop{r}^{2}}-\frac{\Lambda \mathop{r}^{2}}{6}-\frac{2M}{r}.
\label{eq9}
\end{equation}
The graph of this function is also plotted in Fig. (\ref{fig1-1}). \newline
\subsection{Effective Potential}\label{subsec-4}
Before discussing geodesics, let's make a statement about the effective potential energy.
One of the  powerful tools to describe any motion of the free particle in the equatorial plane of a spherically (or cylindrically)
An asymmetric center of attraction is an effective potential method. The effective potential plot instantly shows many central features of particle motion. Indeed, it is a classification scheme to sort orbits and from the properties of the effective potential one can infer several interesting qualitative results about the orbits. For example, circular orbits in the equatorial plane are located at the zeros and the turning points of the effective potential. Indeed, the effective potential plays the same role for geodesic motion as the potential in classical mechanics for one--dimensional motion.

\begin{figure}
\centering
\includegraphics[width=14cm]{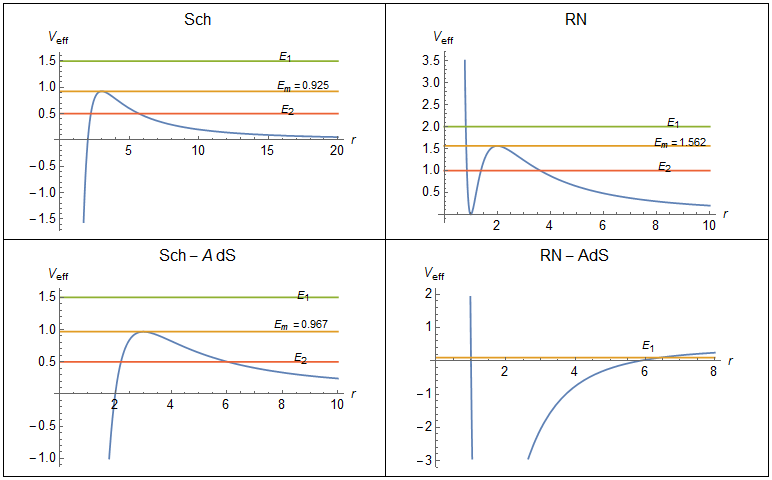}
\caption{Effective potential diagrams for the metric coefficient (1) in linear electrodynamics in the cases of \emph{Sch} (5), \emph{RN} (6), \emph{Sch–(A)dS} (7), and \emph{RN-(A)dS} (9). Numerical values of parameters are: $M=1$, $q=1$ and $\Lambda=-0.9$. The horizontal lines represent the energy levels.}
\label{fig2-2}
\end{figure}

Here is a very brief overview of the effective potential approach to the metric (1). For a more detailed discussion, we can refer to the original texts, e. g. \cite{Airy1}.\\
The effective potential  for massless particle corresponding to metric (1) can be written as \cite{Airy1}:
\begin{equation}
\mathop{V_{eff}}=\mathop{L^2}\frac{f(r)}{\mathop{r}^{2}},
\label{eq10}
\end{equation}
where $L$ is generalized momentum corresponding to the $\phi$ coordinate. By plotting effective potential, the permissible regions of motion and equilibrium (stable or unstable) points can be determined. For this purpose, the graphs of effective potentials corresponding to the background functions (given in  (\ref{eq5}-\ref{eq7}, \ref{eq9}) are plotted in  Fig.\ref{fig2-2}. One can deduce the following by a glance at the figures:\\
1) \emph{Sch} case: For any values of $M$, there is a unstable circular orbit for $r_{m}>r_{h}$\rlap.\footnote{$r_{m}$ and  $r_{h}$  are extremum of effective potential and outer horizon of the \emph{BH}, respectively.}\\
2) \emph{RN} case:  For any values of $M, q$, there is a unstable circular orbit for $r_{m}>r_{h}$.\\
3) \emph{Sch--dS}  case:  For any values of $M, \Lambda$, there is a unstable circular orbit for $r_{m}>r_{h}$.\\
4) \emph{RN--dS} case:  For any values of $M, \Lambda, q$, there is a stable point, but for $\Lambda<0$ stable point is between the inner and outer horizons and thus circular motion is eliminated. For $\Lambda>0$ stable point is outside the horizon and thus circular motion is possible.

\subsection{The Null Geodesics}\label{subsec-3}
\begin{figure}[ht!]
\centering
\includegraphics[width=10cm]{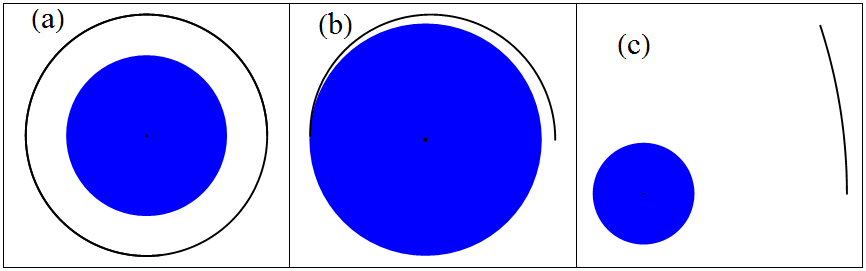}
\caption{The null paths for geodesic equations (12-14) in linear electrodynamics cases when the metric coefficient is given by (5,6,7,9).
Numerical values of parameters are: $M=1$, $q=1$ and $\Lambda=-0.9$.
The initial location ($r_{0}$) and horizon location ($r_{h}$) for each state are as follows and for all three states ($L=5$),\\
a: When the initial location is placed at a suitable distance from the event horizon, the photon path will be circular and the results of \emph{Sch}, \emph{RN} and \emph{Sch–(A)dS} will be qualitatively similar. \emph{Sch} ($\mathop{r}_{h}=2$, $\mathop{r}_{0}=3$), \emph{RN} ($\mathop{r}_{h}=1$, $\mathop{r}_{0}=2$) and \emph{Sch–(A)dS} ($\mathop{r}_{h}\simeq 1.5$, $r_{0}=3$).\\
b: When the initial location is very close to the event horizon, the photon's path is short and eventually absorbed by the black hole. The results of \emph{Sch}, \emph{RN}, \emph{Sch–(A)dS} and \emph{RN-(A)dS} are qualitatively similar. \emph{Sch} ($\mathop{r}_{h}=2$, $\mathop{r}_{0}=2.5$), \emph{RN} ($\mathop{r}_{h}=1$, $\mathop{r}_{0}=1.5$), \emph{Sch–(A)dS} ($\mathop{r}_{h}\simeq 1.5$, $\mathop{r}_{0}=2$) and \emph{RN-(A)dS}: Under any conditions and at any distance, the photon falls into the black hole, ($\mathop{r}_{h}\simeq 13$, $\mathop{r}_{0}=14$).\\
c: When the initial location is far away from the event horizon, the path of the photon gets away from the black hole. The results of \emph{Sch}, \emph{RN} and \emph{Sch–(A)dS} are qualitatively similar. \emph{Sch} ($\mathop{r}_{h}=2$, $\mathop{r}_{0}=10$), \emph{RN}($\mathop{r}_{h}=1$, $\mathop{r}_{0}=10$) and \emph{Sch–(A)dS} ($\mathop{r}_{h}\simeq 1.5$, $\mathop{r}_{0}=10$).\\
(The extent of the \emph{BH} (horizon) is determined by the blue disk).}
\label{fig3-3}
\end{figure}

Let us write down the geodesic equations for background metric (1) when $f(r)$ is given by \ref{eq5}, \ref{eq6}, \ref{eq7}, \ref{eq9}. The geodesic equation can be written as
\begin{equation}
\mathop{{\ddot{x}}^{\alpha }}+\mathop{\Gamma _{\beta \nu }^{\alpha }}\mathop{{\dot{x}}^{\beta }}\mathop{{\dot{x}}^{\nu }}=0
\label{eq11}
\end{equation}
where $\Gamma _{\beta \nu }^{\alpha}$ are affine connections and dots denote derivative concerning a canonical parameter (But not proper time). For motion in the equatorial plane, we set ($\theta =\frac{\pi }{2}$) in spherical and ($z=0$) in cylindrical coordinates. By writing these equations according to the background function, we find:
\begin{equation}
\ddot{t}+\frac{{f}'(r)}{f(r)}\dot{t}\dot{r}=0
\label{eq12}
\end{equation}
\begin{align}
\ddot{r}+\frac{1}{2}f(r){f}'(r)\mathop{{\dot{t}}}^{2}-\frac{1}{2}\frac{{f}'(r)}{f(r)}\mathop{{\dot{r}}}^{2}-rf(r)\mathop{{\dot{\phi }}}^{2}=0
\label{eq13}
\end{align}
and
\begin{equation}
\ddot{\phi }+\frac{2}{r}\dot{r}\dot{\phi }=0.
\label{eq14}
\end{equation}
Recent equations aren't exactly solvable and it is necessary to use numerical methods. In addition, numerical recipes require specific initial conditions to illustrate the geodesic paths. Given the initial conditions as \begin{center}$r_{0}= r_0,\hspace{2mm} \phi(0)= 0, \hspace{2mm} \dot{r}(0)=0$ and $L=\mathop{r}^{2}\dot{\phi}=5.$\end{center}By implementing the numerical recipes, the results (for geodesic paths) are illustrated in Fig.\ref{fig3-3}. More detailed descriptions for Fig.3 are as follows:\\
1) In three cases of background function given by (5-7) (spherical coordinates), choosing $M=q=1$ and $r_{0}=r_{m}>r_{h}$, null geodesic paths are the (unstable) circle(Fig.3-a).\\
2) In \emph{Sch}  case for $M > 1$, and in \emph{RN} case for $M, q > 2$, the paths are qualitatively as (Fig.3-b), that is photo falls in to the \emph{BH}.\\
3) In \emph{Sch} and \emph{RN} cases, for $r_{0}\gg  r_{h}$, the photon path moves away from the \emph{BH} (Fig.3-c).\\
4) In the \emph{$RN-AdS$} case (cylindrical coordinates), for any values of $ M, q, \Lambda$, the paths are qualitative as (Fig.3-b), that is photo falls into the \emph{BH}. It should be noted that the numerical calculations for $\Lambda>0$ have no output.\\

\section{The Effective Metrics in the Non-Linear Electrodynamic}\label{sec-4}
In this section, by taking into account the quadruple term correction to linear Maxwell theory,  a non--linear electrodynamic model in presence of a mimetic field is considered to explore the null geodesic paths of the corresponding spacetime. As mentioned in section 2, to study the geodesic structure of massless particles in nonlinear electrodynamics, an effective metric should be used instead of a background metric. The effective metric is constructed from the background metric and function determined by a particular nonlinear theory considered,  here mimetic theory. Also, as in the previous section, the background metrics are considered in spherical and cylindrical coordinates.

\subsection{Spherically Symmetric Spacetime}\label{subsec-1}
Interested nonlinear  Maxwell theory here is described by the presence of a quadruple term as a correction to the \emph{RN} solution. The corresponding background  metric for this case is specified by the following background function \cite{Airy22}:
\begin{equation}
\mathop{f_{X} (r)}=1-\frac{2M}{r}+\frac{\mathop{q}^{2}}{\mathop{r}^{2}}+\frac{\mathop{q}_{1}^{2}}{\mathop{r}^{4}}    \qquad  ,  \qquad   \omega (r)=\frac{a}{r},
\label{eq15}
\end{equation}
where the quadruple correction is represented by the last term and $q_1$ is the quadruple charge. Also,  the field equations gives  the mimetic field as $\omega (r)=\frac{a}{r}$ \cite{Airy22}. The graph of the function (15) is plotted in Fig.4 (left panel) for the certain values of $M$, $q$ and $q_{1}$. The figures show, that spacetime has a horizon and is asymptotically flat.

\subsection{Cylindrical Spacetime}\label{subsec-2}
The cylindrical spacetime metric (8) in linear electrodynamics is given by the background function (9) which describes an \emph{AdS-RN} spacetime.
In nonlinear Maxwell theory (quadruple case) and the presence of a mimetic field, the field equations give the background  and mimetic functions as follows \cite{Airy22}:
\begin{equation}
\mathop{f_{Y} (r)}=\frac{\mathop{q}^{2}}{\mathop{r}^{2}}-\frac{2M}{r}+\frac{\mathop{\Lambda r}^{2}}{6}+\frac{\mathop{q}_{1}^{2}}{\mathop{r}^{4}}    \qquad  ,  \qquad   \omega (r)=\frac{a}{r},
\label{eq16}
\end{equation}
where the background function includes a quadruple term $q_1^2/r^4$ correction concerning the linear case (9).
The graph of the function (16) is plotted in Fig.4 (right panel) for the certain values of  $M$, $q$, $q_{1}$ and $\Lambda$. The figures show, that spacetime has two horizons as a spherical case and is asymptotically  \emph{(A)dS} one.
\begin{figure}[!pt]
\centering
\includegraphics[width=14cm]{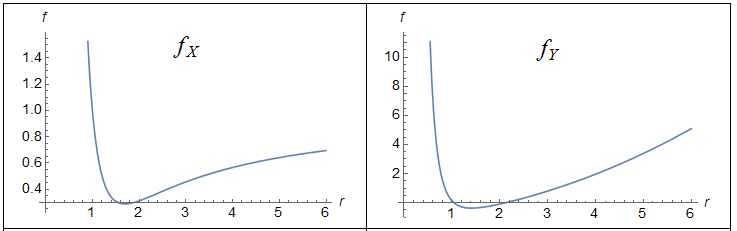}
\caption{The graph of the metric coefficient (17) in nonlinear electrodynamics for the cases $f_{X}$ (15) and $f_{Y}$ (16). Numerical values of parameters are: $M=1$, $q=1$, $\mathop{q}_{1}=1$ and $\Lambda=0.9$.}
\label{fig4-4}
\end{figure}

\subsection{The Null Geodesics of the Effective Metric}\label{subsec-3}
As mentioned, in non--linear electrodynamics, the photon path is given by geodesics of the effective metric. Thus, to obtain the geodetic null paths, we must obtain the effective metric corresponding to the relevant background metric. As can be seen, the effective metric (\ref{eq2}) is constructed by the background function $f(r)$ and the mimetic field $\omega (r)$. Then,  for $\omega (r)=a/r$, it  takes the following form:
\begin{equation}
{ds_{eff}^2}=\sqrt{\frac{a}{r}}f (r)\mathop{dt^2} -\sqrt{\frac{a}{r}}\frac{dr^2}{f (r)}-\sqrt{\frac{r}{a}}\mathop{r^2}\mathop{d\Omega ^2},
\end{equation}
where $f(r)$ is given by (15) and (16). Now, we can proceed to write down the geodesic equations corresponding to the effective metric (17) through equation (11).
As previously, we set ($\theta =\frac{\pi }{2}$) and ($z=0$) in spherically and cylindrically spacetimes, respectively.
Accordingly, for the two spacetimes,  geodesic equations (\ref{eq11}) results in the followings:
\begin{equation}
\ddot{t}+\frac{1}{2}(\frac{2\omega (r){f }'(r)+f (r){\omega }'(r)}{\omega (r)f (r)})\dot{t}\dot{r}=0,
\label{eq17}
\end{equation}

\begin{align}
\ddot{r}&+\frac{f (r)}{4\omega (r)}(2\omega (r){f }'(r)+f (r){\omega}'(r))\mathop{{\dot{t}}^2}-\frac{(2\omega (r){f }'(r)-f (r){\omega }'(r))}{4\omega (r)f (r)}\mathop{{\dot{r}}^2}\nonumber\\
&+\frac{rf (r)}{\mathop{4\omega^2 (r)}}(r{\omega }'(r)-4\omega (r))\mathop{{\dot{\phi }}^2}=0,
\label{eq18}
\end{align}
and
\begin{equation}
\ddot{\phi }+\frac{(4\omega (r)-r{\omega }'(r))}{2r\omega (r)}\dot{r}\dot{\phi }=0.
\label{eq19}
\end{equation}

\begin{figure}
\centering
\includegraphics[width=14cm]{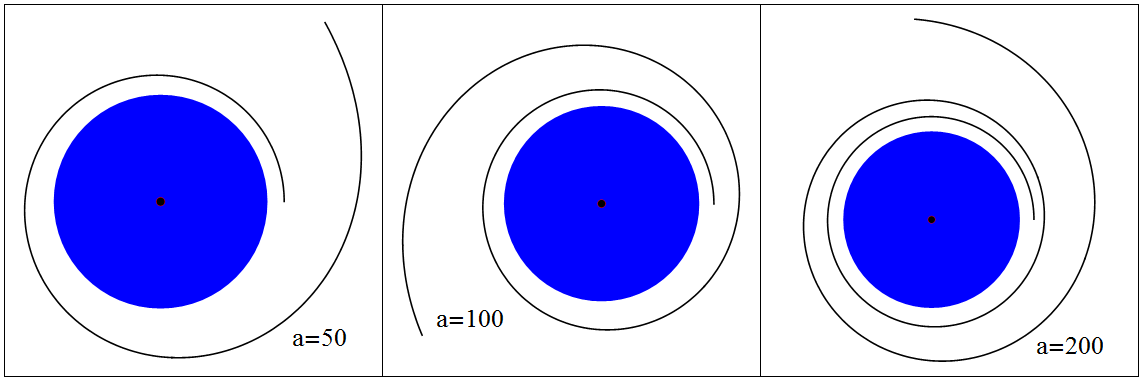}
\caption{The null paths for geodesic equations (18-20) in linear electrodynamics cases when the metric coefficient is given by (15) with different values to $a$ parameter. Numerical values of parameters are: $M=1$, $q=1$ and $\mathop{q}_{1}=1$. (The extent of the \emph{BH} (horizon) is determined by the blue disk).}
\label{fig5-5}
\end{figure}

\begin{figure}
\centering
\includegraphics[width=14cm]{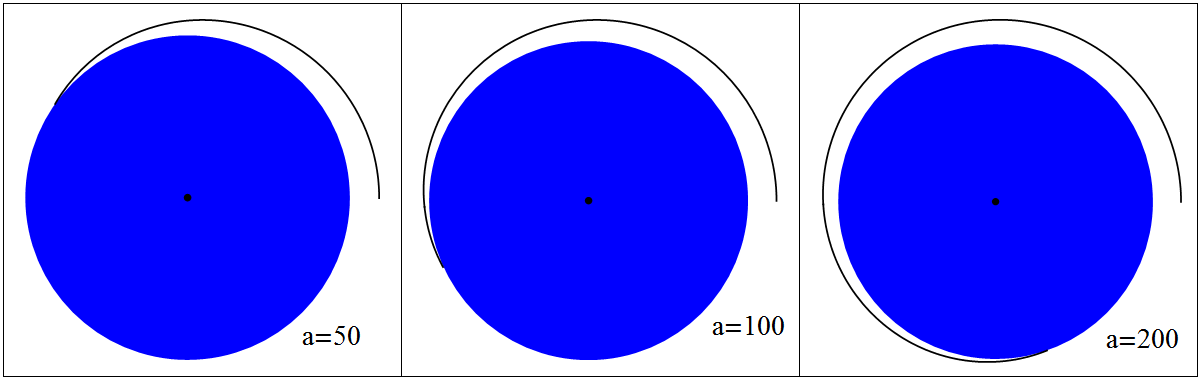}
\caption{The null paths for geodesic equations (18-20) in linear electrodynamics cases when the metric coefficient is given by (16) with different values to $a$ parameter. Numerical values of parameters are: $M=1$, $q=1$, $\mathop{q}_{1}=1$ and $\Lambda=0.9$. (The extent of the \emph{BH} (horizon) is determined by the blue disk).}
\label{fig6-6}
\end{figure}

These equations aren't exactly solvable for both cases of background functions (15), and (16) and it is necessary to use numerical methods. Also, the initial conditions are taken as linear case, that is \begin{center}$r_{0}= r_0,\hspace{2mm} \phi(0)= 0, \hspace{2mm} \dot{r}(0)=0$ and $L=\mathop{r}^{2}\dot{\phi}=5.$\end{center} Finally, the results for the null geodesic paths are illustrated in Fig.\ref{fig5-5} and Fig.\ref{fig6-6}. For better comparison, the values of the parameters ($M,q$) are taken to be the same in the linear and nonlinear cases. By looking at the figures, the following points can be deduced.
\begin{itemize}

\item [1].
Fig.5:\\
Contrary to the linear case (Fig.3-a,b), the photon moves away from the black hole. Notably,  by making the mimetic parameter $a$ larger, the photon moves away later. Physically, it means that, in presence of the stronger mimetic field, the gravity field of the black hole becomes weaker.\\
For $M$, $q$ and $q_{1}>7$, numerical plotting shows the photon gets away from the \emph{BH}, something like the figure Fig.3-c, except that the photon orbits the black hole several times before getting away the horizon.

\item [2].
Fig.6:\\
Before stating the results, it should be noted that numerical calculations show unlike the linear case, where there was no output for $\Lambda>0$, in this case, there is no output for $\Lambda<0$. More precisely, the numerical  recipes  haven't output  graphs for  $\Lambda<0.3$

The results for numerical plotting, in this case, are illustrated in Fig.6. we can deduce two followings:\\
1- Similar to the linear case (Fig.3-b),  the photon falls into the \emph{BH}.  Notably, by making the mimetic parameter $a$ larger, the falling occurs later. In other words, the photon travels longer distances before falling into the black hole. Physically, it means that, in presence of the stronger mimetic field, the gravity field of the \emph{BH} becomes stronger (contrary to the spherical case).\\
2- Note that the spacetime is asymptotically \emph{dS} one ($\Lambda>0$) which has a repulsive character. But as the figure shows, due to the presence of a mimetic field, it acquires an attractive character, and this property is intensified by reducing the mimetic parameter.\\
As a final remark in this section, we note that in the previous section (linear case), some important motion information is obtained using the concept of the effective potential. Especially in the extreme points where circular motions are important. But, in the nonlinear case, it is impossible to definite this concept and therefore circular orbits aren't possible. In essence, this goes back to the Lagrangian formulation. Remember that in the linear case, we can define the effective potential for a spherically symmetric metric when the coefficients of $dt^2$ and $dr^2$ (in metric (1)) are inverses of each other. But, in the nonlinear case, as can be seen from the metric (2) or (17), this isn't the case and therefore the effective potential approach is eliminated. For a better understanding of this issue, we will briefly analyze it. When we use the effective metric to express the Lagrangian in nonlinear electrodynamics, the $\frac{1}{r}$ term appears in the $E$ term, which makes the energy term not constant. Thus, as a physical result, we can say that the presence of the mimetic field doesn't allow the circular orbit around the \emph{BH}.

\end{itemize}

\section{Conclusion}\label{sec-5}
Generally, geodesics are useful tools to explore the spacetime geometry structure. The null geodesics are of especially importance in this regards. In this work, the null geodesics around the mimetic charged black hole coupled to nonlinear electrodynamics are considered. The nonlinear effects appear as a correction to linear theory for the case of \emph{RN} BH in spherical coordinates and \emph{RN--dS} BH in cylindrical coordinates. The important of such model has many astrophysical applications as the main role of the quadruples in generation of gravitational radiation.\\
In nonlinear Maxwell theory, the photon path is followed by the geodesic of the effective metric constructed from the background metric and a function that comes from the considered theory (here mimetic one). Since, the effective metric include the (free) mimetic parameter $a$, it allows to relative manipulations of the photon paths. \\
The notable conclusions can be stated as follows:\\
1) Contrary to the linear case, in a nonlinear regime, there isn't any unstable (or stable) circular orbit.\\
2) Contrary to the pure \emph{RN} case, the presence of quadruple term causes the photon to take a path away from the black hole. The effect of the mimetic parameter is that by making it larger, moving away from the \emph{BH} occurs later.\\
3) Similar to the pure  \emph{RN--dS} case,  the photon falls into the black hole. The effect of the mimetic parameter is that by making it larger, falling into the \emph{BH} occurs later.\\
4) An astrophysical consequence which can be deduced from Fig.6 is that even though the spacetime is asymptotically \emph{dS} one ($\Lambda>0$) and has
a repulsive character, the presence of mimetic field converts this character to the attractive one and this property is intensified by reducing the mimetic parameter.\\

\end{document}